\documentclass[showpacs,preprintnumbers,amsmath,amssymb,openany]{revtex4}

\usepackage{graphicx}
\usepackage{dcolumn}
\usepackage{bm}

\preprint{}

\begin{document}

\title{A new approach for modelling mixed traffic flow with
motorized vehicles and non-motorized vehicles based on cellular automaton model }

\author{Xiao-Mei Zhao }
 \email{xmzhao@bjtu.edu.cn}
\author{Bin Jia}%
 \email{bjia@bjtu.edu.cn}
 \author{Zi-You Gao }%
 \email{zygao@bjtu.edu.cn}

\affiliation{
School of Traffic and Transportation,Beijing Jiaotong University,Beijing 100044,China\\
}%

\date{\today}

\begin{abstract}
In this study, we provide a novel approach for modelling the mixed
traffic flow. The basic idea is to integrate models for
nonmotorized vehicles (nm-vehicles) with models for motorized
vehicles (m-vehicles). Based on the idea, a model for mix traffic
flow is realized in in the following two steps. At a first step,
the models that can be integrated should be chosen. The famous
NaSch cellular automata (NCA) model for m-vehicles and the Burgur
cellular automata (BCA) model for nm-vehicles are used in this
paper, since the two models are similar and comparable. At a
second step, we should study coupling rules between m-vehicles and
nm-vehicles to represent their interaction. Special lane changing
rules are designed for the coupling process. The proposed model is
named as the combined cellular automata (CCA) model. The model is
applied to a typical mixed traffic scenario, where a bus stop
without special stop bay is set on nonmotorized lanes. The
simulation results show that the model can describe both the
interaction between the flow of nm-vehicles and m-vehicles and
their characters.

\end{abstract}

\maketitle

\section{\label{sec:level1}Introduction }
Traffic congestions and the related problems such as traffic
safety problems, environment pollution problems and energy crisis
and so forth are significant for the national economy and the
people's livelihood and commonly exist in most large cities all
over the world. To uncover the traffic nature and clarify the
occurrence of various phenomena in diverse road types, numerous
researchers are devoted to developing traffic flow models. And
substantial progress has been achieved in understanding the origin
of many empirically observed features for roadways with mainly
motorized vehicles(thereafter m-vehicle, including car, bus,
truck) or homogeneous traffic \cite{1,2,2b,3,4}, primarily
reflecting the traffic condition in developed countries. These
achievements help to utilize efficiently limited construction
budget and guide traffic planning and designing, management and
control. The typical example is Lincoln tunnel, where the flow
goes up twenty percent after control. Up to now, considerable
attentions have been focused on traffic flow theory to provide
reasonable advices on alleviating traffic congestion.

However, in developing countries, e.g. China, India, Bangladesh
and Indonesia \cite{2b}, m-vehicles come in increasing numbers,
and simultaneously nonmotorized vehicles(thereafter nm-vehicle,
including bicycle, three-wheeler, motorcycle) are still prevalent
for most short-distance trips due to low income levels or
convenient parking. Thus, m-vehicles and nm-vehicles always blend
on roads without isolations between motorized lanes(m-lane) and
non-motorized lanes(nm-lane) or intersections. The mix traffic or
heterogeneous traffic with both m-vehicles and nm-vehicles will
persist for further years, since some governments advocate that
citizens take bicycles for a short distance instead of driving
cars to release issues on lack of energy and environmental
pollution. It is noted that the mix or heterogeneous traffic in
the following text means traffic with the mixture of m-vehicles
and nm-vehicles. The chief difference between m-vehicles and
nm-vehicles is that behaviors of m-vehicles are lane-based, while
nm-vehicles do not follow each other within lanes but move in both
longitudinal and lateral direction\cite{2b}. The prominent
characters of nm-vehicles are much flexible, low-speed and
unsubstantial. When two kinds of vehicles mix somewhere,
m-vehicles should concede nm-vehicles to guarantee the security of
drivers of nm-vehicles. Apparently, the mix traffic flow would be
much more complicated than the homogeneous flow. In motorized
traffic flow theory, there are mainly two kinds of microscopic
traffic models, cellular automaton(CA) models and
car-following(CF) models. It is reported that few devotes have
been done on expanding these models into investigating the problem
of the mixed traffic flow. Inhomogeneous CA models based on
non-identical particle size are presented to characterize the
behaviors of vehicular movements in a mixed traffic environments
with various motorized vehicles \cite{4}. The model applicable to
the cases of car-bicycle following are investigated by Faghri
\cite{5}. Oketch \cite{6} incorporated car-following rules and
lateral movement to model mixed-traffic flow. Wu and Dai et al.
\cite{7b} introduced a CA model for mix traffic flow with
m-vehicles and motorcycles. However, either CA models or CF models
can not be suitable to exhibit both lane-based behaviors of
m-vehicles and non-lane-based bahaviors of nm-vehicles. Cho and Wu
\cite{7} proposed a model of motorcycles with longitudinal and
lateral movement, and pointed out that this model could be the
basis of bicycle or pedestrian flow model.Obviously, it is
inappropriate to directly extend motorized traffic flow theory
into mixed traffic systems, because complicate interferences
between m-vehicles and nm-vehicles can not be rightly described in
the models for m-vehicles. It is a long-standing tradition to
neglect the mix traffic mode that represents the status of
transportation in developing countries. To get deep insights into
the mixed traffic flow, it is important to develop appropriate
models describing the general feature of mix traffic flow and
disclose the basic discipline, furthermore to enhance transporting
efficiency in mix traffic systems and provide the firm
infrastructure for the sustainable growth of national economics.

In the previous studies, many works about mix traffic flow models
were done on extending only one kind of motorized traffic models
into describing the characters of heterogeneous vehicles. We think
it is more reasonable for mix traffic flow models to integrate
models for nonmotorized transportation modes with models for
motorized modes. Two key problems should be solved. One is how to
choose suitably models that may be comparable, the other is how to
establish the relationship among different types of models. As for
the former, CA models may be good choices for m-vehices, owing to
the relatively simple rules in expressions and better descriptions
of most realistic traffic phenomena. At the same time, multi-value
CA (MCA) models developed by Nishinari and Takahashi \cite{8,9,10}
can depict the multi-lane traffic without explicitly considering
the lane-changing rule, and can be used to describe the
non-motorized traffic flow. CA and MCA models are similar in the
following two aspects. Both two models are discrete in the time
and space, and states of vehicles are updated related rules in
forwards motion. So The two models can be perfectly comparable. To
solve the latter, identical dimensions of sites are used in
different kind of CA models. In this paper, the aim is to
establish a novel approach for modelling mix traffic flow based on
the combination of the CA model for motorized traffic flow and the
MCA model for non-motorized one. So it is referred as the combined
CA (CCA) model.The model is applied to simulate a special mixed
system, where the bus stop inserted into the nm-lane, and
nm-vehicles and m-vehicles mix near the stop. The simulation
results indicate that the CCA model can not only correctly
character both nonmotorized and motorized transportation modes but
also properly display their interactions. Thus, it is reasonable
to depict the chief properties of mix-traffic flow.

In this paper, we choose the typical NaSch cellular automaton
(NCA) model \cite{3} for the CA model and the Burgers cellular
automaton (BCA) model \cite{8,9,10,11,11b} for the MCA model.
Furthermore, other improved CA models and MCA models or other
suitable traffic flow models can be used in the proposed approach.
The approach also can be generalized to other cases of the mixed
traffic systems, such as the intersection, the roads without the
isolation between the motorized lane(m-lane) and the nonmotorized
lane(nm-lane) et al.

The remaining parts of the paper is organized as follows. The
mixed traffic system is introduced and the CCA models is presented
in section 2. Section 3 presents the simulation results and the
discussions. Finally, the summary and the further studies are
addressed.

\section{\label{sec:level1}The combined CA models(CCA models)}

The basic idea of the approach for modelling mix traffic flow is
to unify MCA models for m-vehicles with BCA models for
nm-vehicles. The main issues are to pick up models for two
transportation modes and connect them to reflect interactions
between two kinds of vehicles. Since the CA and the MCA model can
reproduce basic phenomena of m-vehicles and nm-vehicles
respectively and have the similar manner, it is expected that the
two models can better be combined and their combination can
exhibit the characteristics of both two kinds of vehicles. And
special lane-changing rules between m-vehicles and nm-vehicles are
designed to build up connections between two kinds of models and
interactions among vehicles. Then we present a simple model to
characterize the traffic flow with the mixture of motorized and
non-motorized vehicles. The new model, combined the NCA and the
BCA model, is named as combined CA (CCA) model.

\subsection{\label{sec:level2}The mixed traffic system}
Bus stops are essential infrastructures for public transport
systems. In undeveloped countries, most bus stops have no special
stop bay, and are set on nm-lanes. Thus, near these bus stops,
buses occupy the nm-lane and block lots of nm-vehicles. According
to rules for nm-vehicles in China \cite{11c}, nm-vehicles is only
permitted to run on nm-lanes, however nm-vehicles can employ the
neighboring m-lane under the guarantee of safety when nm-vehicles
on nm-lanes are blocked by hindrances. Thus some nm-vehicles would
rush into the adjacent m-lane if buses dwells at the stop on
nm-lanes. The case mentioned above is a typical example for the
mixed traffic flow. Here, the mixed traffic system near these bus
stops will be investigated. Consider the mixed traffic system with
two lanes including a nm-lane and a m-lane. The traffic system is
sketched in Fig. 1. The road configuration with two lanes, is
split into five sections, section A, B, C, D and E. Section A and
E are the entrance and exit region of the road, respectively.
Section C on the nm-lane is the bus stop. Section B and D on the
nm-lane are the upstream part and downstream part of the bus stop,
respectively.

Each lane is divided into L sites with identical size, which are
named as NCA sites for nm-vehicles and BCA sites  and for
m-vehicles, respectively. The uniform dimension of different sites
can simplify the computation process. Assume that each BCA site
can hold M nm-vehicles at most, and each NCA site may be either
empty or occupied by one m-vehicles. Movement of vehicles includes
forward motions and lane-changing motions. M-vehicles in NCA sites
move forwards according to the evolution rules of the NCA model,
and nm-vehicles in BCA sites according to the rules of the BCA
model. The speeds of all vehicles are integer values. The maximal
speed of m-vehicles(nm-vehicles) is $v^m_{max}$($v^{nm}_{max}$).
In the following, the variables with the superscript $m$($nm$)
denote those of m-vehicles(nm-vehicles).

The system contains two types of m-vehicles and one type of
nm-vehicles. The m-vehicles have cars with one-site length and
buses with two-sites length. The mixing probability $P_m$ stands
the proportion of buses in m-vehicles. All buses must halt at the
bus stop, and is called as stopping buses. $T_s$ denotes the dwell
time of stoping buses at the stop. After the dwelling procedure,
stopping buses is regarded as non-stop buses.

\subsection{\label{sec:level2}The rules of the NCA models}

The NCA model is typically employed to control the forward motion
of m-vehicles. At each discrete time step $t\rightarrow{t+1}$, the
state of each vehicle is updated by the following rules \cite{3}:
\begin{enumerate}
    \item Acceleration, $v_j(t+1/3)\rightarrow\min{\{v_j(t)+1,v^m_{max}\}}$;
    \item Deceleration, $v_j(t+2/3)\rightarrow\min{\{v_j(t+1/3),d_j(t)\}}$;
    \item Randomization, $v_j(t+1)\rightarrow\max{\{v_j(t+2/3)-1,0\}}$ with probability $p$;
    \item Motion, $x_j(t+1)\rightarrow{{x_j(t)}+{v_j(t+1)}}$.
\end{enumerate}
Here, $x_j(t)$ and $v_j(t)$ are the head position and velocity of
m-vehicle $j$ in time step $t$, $d_j(t)$ is the number of empty
sites between vehicle $j$ and its nearest preceding site occupied
by vehicles. $p$ is the randomization probability in time step
$t$. For simplicity, only the determined NCA model with $p=0$ is
used in the following simulations.

If vehicle $j$ is the nearest vehicle behind the bus stop and a
stopping bus , the gap $d_j(t)$ is computed as
$d_j(t)=x_{C}-x_j(t)$, where $x_{C}$ is the rightmost position of
Section C, namely the end of the bus stop.

For the stopping bus $j$ at the bus stop, if its dwelling time
$T_{s,j}(t)$ is less than $T_s$, then it continues to halt at the
stop, and the dwelling time is updated as
$T_{s,j}(t+1)=T_{s,j}(t)+1$. Otherwise, the bus becomes a non-stop
bus.

\subsection{\label{sec:level2}The rules of the BCA models}

In the BCA model, the lane-changing rule is neglected. For
simplicity the maximal speed $v^{nm}_{max}$ is set at $1$. Of
course, other values for $v^{nm}_{max}>1$ can also be considered,
and the complexity of the computation increases. The number of
vehicles in each site evolves as follows \cite{8,9,10,11}
\begin{equation}\label{e1}
    U_j(t+1)=U_j(t)-\min{(U_{j}(t),D_{j-1}(t))+\min{(U_{j+1}(t),D_j(t))}}
\end{equation}
where $U_j(t)$ represents the number of nm-vehicles at site $j$
and time $t$, $D_j(t)=M-U_j(t)$.

If the site $j-1$ in front of the current site $j$ is occupied by
m-vehicles at time $t$, then $D_{j}(t)=0$.

\subsection{\label{sec:level2}Lane-changing motions}
Only nm-vehicles and buses near the bus stop are permitted to
change lanes. Buses in section B and C will change lanes
asymmetrically. For simplicity, the lane-changing rules similar to
those for off-ramp traffic systems in ref.[12] are used in this
paper. This is because the lane-changing behaviors of vehicles on
the main road, which leave the main road and enter the off-ramp,
are similar to those of buses entering and leaving the bus stop.
The drivers of buses are willing to run on the nm-lane to stop
conveniently when they are close to the bus stop. These buses will
change from the left lane to the right lane as long as conditions
on the right lane are not worse than those on the left lane.
Namely, if the following condition is satisfied\cite{12},
\begin{eqnarray}
&&[d_j=d_{j,other}=0\ \mbox{ or }\ ( d_{j,other}\neq 0 \nonumber\\
&& \mbox{and}\ d_j-d_{j,other}\leq 2 )]\ \mbox{ and }\
d_{j,back}>v_{ob},
\end{eqnarray}
the stopping bus will change from the m-lane to the nm-lane with
the probability $P_{c1}$ in sections B and C.
$d_{j,other}(d_{j,back})$ represents the number of empty sites
between vehicle $j$ and its nearest preceding (back) unempty site
on the destination lane at time $t$. $v_{ob}$ represents the
velocity of its back vehicle on the destination lane, and
$v_{ob}=v^{nm}_{max}$ when its nearest following vehicles are
nm-vehicles. Condition $d_j=d_{j,other}=0$ means that there is no
gap to move forward on both lanes in the next time step; Condition
$d_j-d_{j,other}\leq 2$ means that the road situation on the
present lane is not much better than that on its neighbor. If a
stopping bus cannot change to the destination lane, until it
approaches $x_{C}$, it would stop to wait for the change chance
(e.g. it will change the lane as soon as the corresponding
position on its right-side lane is empty). In sections B and C,
stopping buses are prohibited changing from the nm-lane to the
m-lane. The bus that has finished the dwelling procedure will
become a non-stop bus. The same lane-changing rules in Eq.(2) are
used for lane-changes of non-stop buses from the nm-lane to the
m-lane in sections C and D. Non-stop buses in sections C and D are
forbidden changing to the nm-lane. If non-stop buses on the
nm-lane still cannot change to the m-lane, when it reaches the
rightmost position of section D $x_{D}$, it would stop to wait for
the change chance mentioned above. Cars are forbidden running on
the nm-lane.

In sections B and C, nm-vehicles in the BCA site $j$ may change
from the nm-lane to the m-lane under the hindrance of the
preceding m-vehicles, if the following conditions are fulfilled,
\begin{eqnarray}
&& d_j<v^{nm}_{max} \ \mbox{ and } \ d_{j,other}>d_j\nonumber\\
&& \ \ \ \ \ \ \ \ \ \mbox{ and }\ d_{j,back}>d_{safe}
\end{eqnarray}
all nm-vehicles in the current BCA site change to the m-lane with
the probability $P_{c1}$. $d_{safe}$ is a safety distance to avoid
crash, and is set to the maximal velocity of its back vehicle on
the destination lane.

Under the situation that some nm-vehicles are running on the
m-lane just in front of the current BCA site $j$, the possibility
of the lane-changing behaviors of present nm-vehicles in site $j$
from the nm-lane to the m-lane may be increased, as long as the
corresponding site on the m-lane still has space to hold
nm-vehicles. Because people would act in conformity with the
majority. Namely, if the following condition are met
\begin{eqnarray}
&& U_{j,other}<M
\end{eqnarray}
then no more than $M-U_{j,other}$ nm-vehicles in BCA site $j$ can
change from the nm-lane to the m-lane with the probability
$P_{c3}$. $U_{j,other}$ represents the number of nm-vehicles in
the corresponding site of the destination lane at time $t$.
Generally, $P_{c3}>P_{c2}$. This means that nm-vehicles would
prefer to change to the m-lane in this case than other cases. The
same changing rules in Eq.(3-4) are used for lane-changes from the
m-lane to the nm-lane in sections C and D. If nm-vehicles cannot
change from the m-lane to the nm-lane, when they approach $x_{D}$,
all or part of them will change to the nm-lane as soon as the
corresponding position on its destination lane isn't completely
occupied.

In addition, to guarantee the avoidance of complete congestion
around the bus stop, the nm-vehicles of the site $x_C-2$ on the
nm-lane will give way to stopping buses with the probability
$P_b$.

\subsection{\label{sec:level2}Boundary conditions}
The simulations are carried out under open boundary condition. In
each time step, when the update of m-vehicles on the road is
finished, we check the positions $x^m_{last}$ of the last
m-vehicles on the entrance of the m-lane. If
$x^m_{last}>v^m_{max}$, a m-vehicle with velocity $v^m_{max}$ is
injected with the inflow rate $P^m_e$ at the site
$\min{\{v^m_{max},x^m_{last}-v^m_{max}\}}$. On the nm-lane, if the
first site isn't full filled with nm-vehicles, nm-vehicles are
inserted with the probability (inflow rate) $P^{nm}_e$ at the
first site. $M$ times of circulation will be done in each time
step. In each circulation, a nm-vehicle will be added on the first
site with the probability $P^{nm}_e$, if there is space on the
first site. The leading vehicles on each lane go out of the system
at $L$ and its following vehicle becomes the new leader.

\section{\label{sec:level1}Simulation results and discussions }
In this section, the characteristics of mixed traffic flow is
discussed in the traffic system mentioned above. Let us consider
the road with $L=500$ sites, the lengthes of sections A, B, C, D
and E are set as $L_A=250$, $L_B=3$, $L_C=4$, $L_D=3$ and
$L_E=240$ sites, respectively. Each site corresponds to 7.5 $m$,
and each time step corresponds to 1 $s$. The model parameters are
set as follows: $M=6$, $v_{max}^{car}=3$, $v_{max}^{bus}=2$,
$P_{c1}=P_{c2}=0.6$, $P_{c3}=1.0$, where the superscript
$car$($bus$) denotes the parameter of the car (bus). According to
the Transit Co-operative Research Program (TCRP) Report 19 (1996)
\cite{13}, the average peak-period dwell time exceeds $30$s per
bus, so we set $T_s=30$s. The mix probability $P_m$ is $0.15$.

\subsection{\label{sec:level2}Phase transition}
Figs.2-3 display the relationships between the flow and the inflow
rate in the CCA model for m-vehicles and nm-vehicles in the case
of $P_b=0.05$, respectively. $q_m$ and $q_{nm}$ represent the flow
of the m-vehicles and the nm-vehicles, respectively. Five virtual
detectors are fixed on site $100$, $200$, $300$, $400$ and $500$
of the road, where the numbers of nm-vehicles and m-vehicles
passing through are recorded. $q_m$ is the average number of
m-vehicles passing through five virtual detectors in each time
step. $q_{nm}$ is the average value that the number of penetrating
nm-vehicles divided by $M$. The first 50,000 time steps are
discarded to avoid the transient behaviors. The flow is averaged
by 100,000 time steps. From Fig. 2(a)(Fig. 3(a)), we find that a
critical inflow rate $P^{m}_{ec}$ ($P^{nm}_{ec}$)(which is pointed
out in Fig.2(a)(Fig.3(a)) only for $P^{nm}_e=1$ ($P^{m}_e=1$))
divides the flow into two regions, the free-flow one and the
saturated-flow one.
\begin{enumerate}
    \item In the region of $P^{m}_e<P^{m}_{ec}$ ($P^{nm}_e<P^{nm}_{ec}$),
     the flow of m-vehicles (nm-vehicles) is free and
    $q_m$ ($q_{nm}$) only depends on itself inflow rate $P^{m}_e$
($P^{nm}_e$).

    \item In the region $P^{m}_e\geq P^{m}_{ec}$ ($P^{nm}_e\geq P^{nm}_{ec}$),
the flow of m-vehicles (nm-vehicles) is saturated. The flow $q_m$
($q_{nm}$) is independent of itself rate $P^{m}_e$ ($P^{nm}_e$),
and reaches its saturation value $q_{mc}$ ($q_{nmc}$). However,
with the increase of $P^{nm}_e$ ($P^{m}_e$), both the flow
$q_{mc}$ and the critical value $P^{m}_{ec}$ ($q_{nmc}$ and
$P^{nm}_{ec}$) reduce until it reaches a minimum. This also can be
obviously observed in Fig. 2(b)(Fig. 3(b)), where the flow versus
itself inflow rate $P^{m}_e$ ($P^{nm}_e$) in the cases of
different $P^{nm}_e$ ($P^{m}_e$) is displayed. The saturated value
$q_{mc}$ and the critical value $P^{m}_{ec}$ ($q_{nmc}$ and
$P^{nm}_{ec}$) decline from $0.34$ and $0.3$ ($0.5$ and $0.58$) at
$P^{nm}_e=0$ ($P^{m}_e=0$) to $0.15$ and $0.1$ ($0.19$ and $0.28$)
at $P^{nm}_e=0.2$ ($P^{m}_e=0.14$). The drop ratios of $q_{mc}$
and $q_{nmc}$ are about $56$ percent and $62$ percent. This
suggests that the mixture of the nm-vehicles and m-vehicles has a
negative effect on the saturated flow of two flows which descends
in a wide range.
\end{enumerate}
Thus, in the proposed CCA model, the phase transition from free
flow to the saturation for both two flows is observed. And the
flow $q_m$ ($q_{nm}$) relies on not only itself inflow rate
$P^{m}_e$ ($P^{nm}_e$), but also $P^{nm}_e$ ($P^{m}_e$). The
mixture of the nm-vehicles and m-vehicles in the traffic system
results in the drop of the saturated flows. Therefore, the model
can exhibit the interactions between nm-vehicles and m-vehicles in
the mixed traffic system.

It is interesting that the collective effect of the nm-vehicles
and m-vehicles only appears when $P^m_e$ or $P^{nm}_e$ surpasses
its critical value. According to the two critical values, the
phase diagram in $(P^m_e,P^{nm}_e)$ space presented in Fig. 4(a)
is classified into four regions, where the flow $q_m$ and $q_{nm}$
are related to $P^m_e$ or $P^{nm}_e$ or both of them. In Regions
$I$ and $II$, the m-flow is free, while it becomes saturated in
Regions $III$ and $IV$. In Regions $I$ and $III$, the nm-flow is
in the state of free flow, while reaches its saturation in Regions
$II$ and $IV$. $line_1$($line_4$) is the boundary of Regions $I$
($II$) and $III$ ($IV$), which corresponds to the critical point
$P^m_{ec}$. It can be found that $P^{m}_{ec}$ decreases firstly
and then maintains at a constant with the increase of the inflow
rate $P^{nm}_{e}$ of the nm-vehicles flow. This indicates that the
mutual effect between the two flows grows gradually and becomes
saturated at the cross point $O:(P^{nm}_{eo},P^{m}_{eo})$. The
same can be observed in the curve of the critical value
$P^{nm}_{ec}$ ($line_2$ and $line_3$).

To get a deep insight into these regions, space-time plots are
depicted in Fig. 5. The left correspond to those of the m-lane,
and the right correspond to the nm-lane. Blue points and green
points represent cars and buses on NCA sites. Red points, black
points and magenta points denote BCA sites with 1-2, 3-4, and 5-6
nm-vehicles. Here, no time-space plots in Regions is shown to
decrease the size of the manuscript. The plots can be provided if
readers send an email to me.
\begin{enumerate}
    \item In Region $I$ , the traffic flow on both two lanes is free flow,
the $q_{m}$ and the $q_{nm}$ depend on only itself inflow rate.
    \item Region $II$, where the $q_m$ only varies with the $P^{m}_e$,
and the $q_{nm}$ gets saturated, the saturation only depends on
the $P^m_e$. The $P^m_e$ is very low, thus m-vehicles are sparse
on the road. Although the lane changing behavior of nm-vehicles
from the nm-lane to the m-lane occurs and a short waiting queue
forms upstream these nm-vehicles, the queue in the m-lane
disappears within several time-steps. Thus, the flow of m-vehicles
won't be perturbed by these nm-vehicles. As the $P^{nm}_e$
increases, nm-vehicles on the road become denser. Most NCA sites
in the nm-lane are fully filled with nm-vehicles. So buses halting
at the stop hinder the forward motion of nm-vehicles, and a long
waiting-queue stretches far from the position of buses, leading to
the reduce of $q_{nmc}$.
    \item Region $III$, where the flux $q_{nm}$ is only dependant
on the $P^{nm}_e$, and the $q_m$ reaches its saturated value
$q_{mc}$, and the saturation flow decreases with $P^{nm}_e$. The
situation is just contrary to that of region $II$. Some
nm-vehicles in the nm-lane accumulate behind the buses, and
dissolve soon after the buses start to run due to low $P^{nm}_e$.
But the lane-changing behaviors of these nm-vehicles strongly
interrupt the movement of m-vehicles in the m-lane, and cause the
reduce of $q_{mc}$.
    \item Region $IV$, where both $q_m$ and $q_{nm}$ remain
constant.
\end{enumerate}

\subsection{\label{sec:level2}The total flow of the system}
To measure the total flow of the mixed traffic system, a total
flow ratio $R_q$ is defined as follows
\begin{eqnarray}
&&R_q=\frac{q_m}{q_{mc0}}+\frac{q_{nm}}{q_{nmc0}}
\end{eqnarray}
where $q_{mc0}$($q_{nmc0}$) is the saturation flow at $P^{nm}_e=0$
($P^{m}_e=0$). Fig.5(a) shows $R_q$ versus $P^{m}_e$ and
$P^{nm}_e$. It can be seen that $R_q$ depends on both two inflow
rates, and the corresponding phase diagram in $(P^m_e,P^{nm}_e)$
space (see Fig. 4(b)) also is divided into four regions similar to
Fig. 4(a).
\begin{enumerate}
\item In Region $I$, both two flows are free, thus the $R_q$
linearly increases with the $P_m^e$ and the $P_{nm}^e$.
 \item In
Region $II$, since the nm-flow is saturated, $R_q$ just varies
with $P^m_e$. Furthermore, Region $II$ can be separated into two
parts $II_1$ and $II_2$, where $R_q$ is independent on $P^{nm}_e$.
In $II_1$, $R_q$ increases with $P^{m}_e$, while $R_q$ decreases
with $P^{m}_e$ in $II_2$. For the convenience of analysis and
comparison, $R_q$ and the corresponding lane-changing times of
nm-vehicles with $P^m_e$ are shown in Fig. 5(b-c) to investigate
the feature of mixed flows. It is can be found that the
lane-changing times of the nm-vehicles increase with the $P^{m}_e$
in $II_1$, most of nm-vehicles can pass through the bus stop by
utilizing the m-lane, and won't hinder m-vehicles due to large
headway on the m-lane. Therefore, $R_q$ increases with the $P^m_e$
as the lane-changing times of the nm-vehicles increase. Whereas
with the further increase of $P^{m}_e$, the gap between
neighboring vehicles gets smaller and less nm-vehicles can succeed
in changing to the m-lane, thus $R_q$ drops with $P^{m}_e$ in
$II_2$.
 \item In Region
$III$, the m-flow is saturated, thus $R_q$ only relies on
$P^{nm}_e$. Also, Region $III$ in Fig. 4(a) can be separated into
two parts $III_1$ and $III_2$,, where $R_q$ is independent on
$P^{m}_e$. In $III_1$, $R_q$ decreases with $P^{nm}_e$, while
$R_q$ increases with $P^{nm}_e$ in $III_2$. Fig. 5(d-e) show $R_q$
and the corresponding lane-changing times of nm-vehicles in Region
$III$. Since the flow of m-vehicles reaches the saturation, the
increase of nm-vehicles changing to the m-lane will cause the
great drop of $q_m$, which is more than the increase of $q_{nm}$
with $P^{nm}_e$. Thus $R_q$ will decrease with $P^{nm}_e$ in
$III_1$. With the further increase of $P^{nm}_e$, the decrease of
lane-changing times of nm-vehicles make that the drop of $q_m$ is
less than the increment of the $q_{nm}$. So $R_q$ starts to
increase with $P^{nm}_e$ in $III_2$.
 \item In Region
$IV$, both two flows reach the saturation values, thus the $R_q$
is independent on both two inflow rate and keeps a constant.
\end{enumerate}

From the discussions above, it can be concluded that lane-changing
behaviors of nm-vehicles are helpful to the  total flow of the
traffic system, when the flow of m-vehicles is free. Especially at
the boundary between $II_1$ and $II_2$, $R_q$ approaches a local
maximum. Contrarily, lane-changing behaviors of nm-vehicles are
harmful to the  total flow, when the flow of m-vehicles is
saturated. At the boundary between $III_1$ and $III_2$, $R_q$
approaches a local minimum.

\section{\label{sec:level1}Conclusions }

A combined cellular automaton (CCA) model is presented to describe
the mixed traffic system composed of m-vehicles and nm-vehicles.
The CCA model is based on the NaSch CA (NCA) model for m-vehicles
and the Burgers CA (BCA) model for nm-vehicles. In the CCA model,
there are two types of sites with identical size, NCA sites and
BCA sites. A NCA site is defined as the site occupied by a
m-vehicle, and updates according to the rules of NCA model. A BCA
site contains a number of nm-vehicles, and its state evolves
according to the rules of BCA model. Thus, the new model is
convenient to perform on the computer. The model is applied to the
mixed traffic system near the bus stop without the special stop
bay, and special lane-changing rules are employed. Firstly, for
the nm-vehicles(m-vehicles) flow, the phase transition from free
flow to saturated flow can be observed at the critical value
$P^m_{ec}$($P^{nm}_{ec}$). According to the two critical values,
the phase diagram in ($P^{nm}_e$,$P^m_e$) space is categorized
into four regions, including Region $I$(where both two flows are
free), Region $II$ (where the flow of nm-vehicles is saturated,
and that of m-vehicles is free), Region $III$ (where the flow of
m-vehicles is saturated, and that of nm-vehicles is free), and
Region $IV$(where both flows reach the saturations). Secondly, to
measure the total flow of the mixed traffic system, a total flow
ratio $R_q$ is introduced. According to the characteristics of
$R_q$,  Region $II$ and $III$ in the space of ($P^{nm}_e$,$P^m_e$)
can be separated into two parts further, where the $R_q$ has a
increasing or decreasing tendency due to the mixture of two flows,
respectively. From these, it can be inferred that the proposed CCA
model could reflect feature of mixed traffic flow very well, and
has great potentials on the practical application. It is noted
that to improve the proposed model for mix traffic system and
validating experimentally it, empirical data investigation and
related calibration are in progress.

The work is only the first step towards understanding characters
of mixed traffic flow.  There are numerous aspects that require
further investigation, such as how to apply the proposed method in
intersections and other traffic conditions to model mix traffic
flow, how to consider the pin-effects of nm-vehicles and
differences among various types of vehicles, how to improve
negative effects induced by mixture of nm-vehicles and m-vehicles
etc. We are planning to address these issues in our future work.

\begin{acknowledgments}
This paper is financially supported by 973 Program (2006CB705500),
Project (70631001 and 70501004) of the National Natural Science
Foundation of China, and Program for Changjiang Scholars and
Innovative Research Team in University(IRT0605).
\end{acknowledgments}

\newpage{}
\textbf{Figures}\\

{\textbf{Fig. 1} The sketch of the road in the mixed traffic
system.}\\

{\textbf{Fig. 2} (a) The variation of the flow $q_m$ of m-vehicles
with the entering probability $P^{nm}_e$ and $P^m_e$. (b) the flow
$q_m$ varies as the $P^m_e$ at fixed $P^{nm}_e=0,0.1,0.2,1$.
$P_b=0.05$. The $P^{nm}_e$ varies from $0$
to $1$ with identical interval $0.02$ in the $P^{nm}_e$-axis.}\\

{\textbf{Fig. 3} (a) The variation of the flow $q_{nm}$ of
nm-vehicles with the entering probability $P^{nm}_e$ and $P^m_e$.
(b) the flow $q_{nm}$ varies as the $P^{nm}_e$ at fixed
$P^m_e=0,0.1,0.14,1$. $P_b=0.05$. The $P^{m}_e$ varies from $0$ to
$1$ with identical interval $0.02$ in the
$P^{m}_e$-axis.}\\

{\textbf{Fig. 4} (a) The phase diagram $(P^m_e,P^{nm}_e)$ in the
mixed traffic system. (b) The redrawn phase diagram $(P^m_e,P^{nm}_e)$ obtained by the variation of the $R_q$.$P_b=0.05$.}\\

{\textbf{Fig. 5} The total flow rate $R_q$(a) versus the inflow
rate $P^{nm}_e$ and $P^m_e$, (b-c) the $R_q$ and the corresponding
lane changing times of nm-vehicles versus $P^m_e$ at
$P^{nm}_e=0.6$ in Region $II$,(d-e) the $R_q$ and the
corresponding lane changing times of nm-vehicles $P^{nm}_e$ at $P^{m}_e=0.6$ in Region $III$.$P_b=0.05$.}\\

\newpage{}
\begin{figure}
\includegraphics[width=5 in]{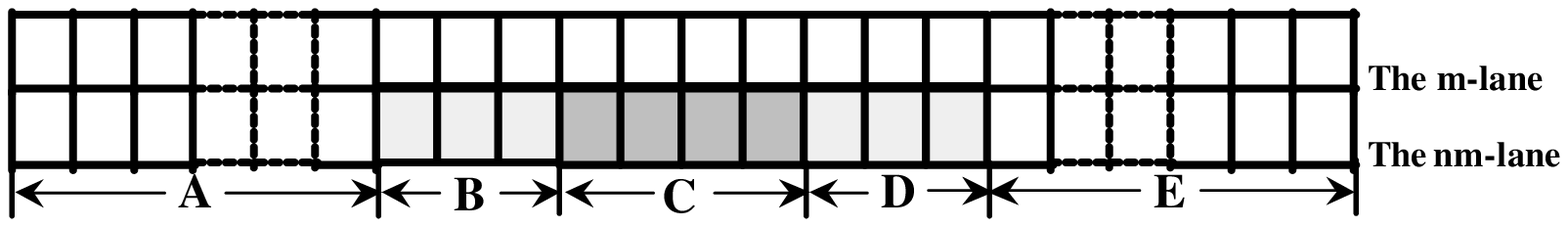}
  \caption{}\label{f1}
\end{figure}

\begin{figure}
\includegraphics[width=5 in]{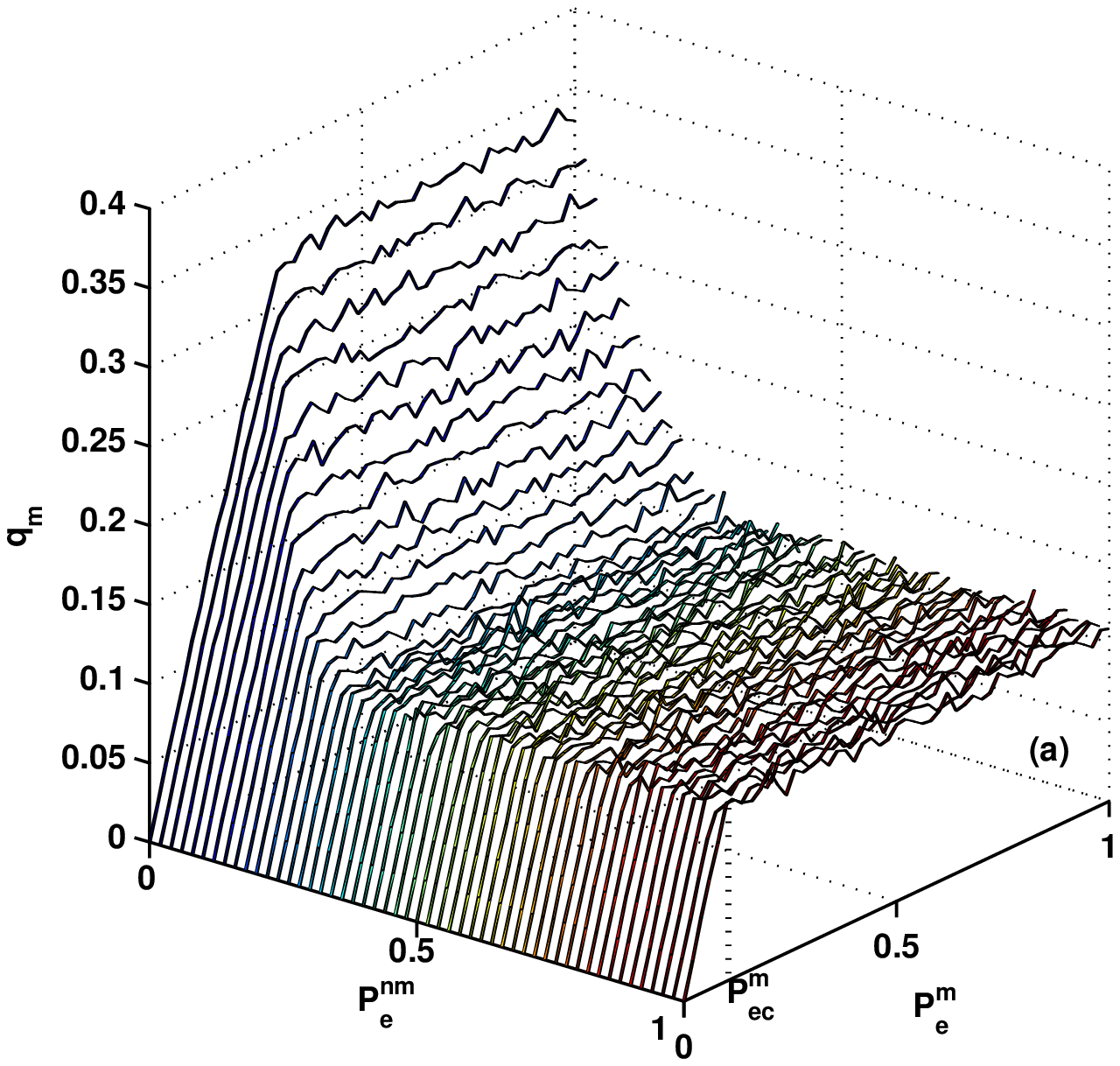}
\includegraphics[width=5 in]{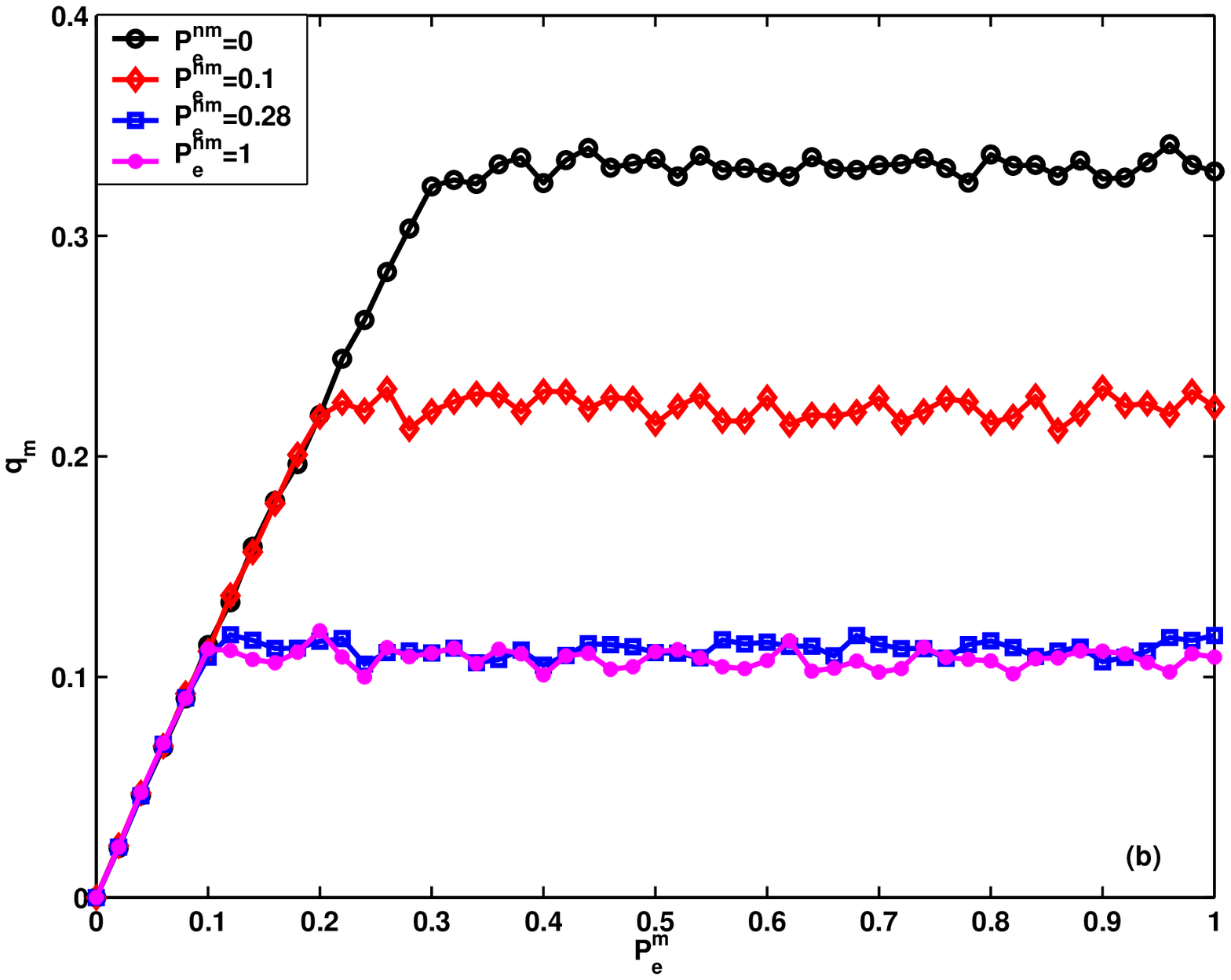}
  \caption{}\label{f2}
\end{figure}

\newpage{}
\begin{figure}
\includegraphics[width=5 in]{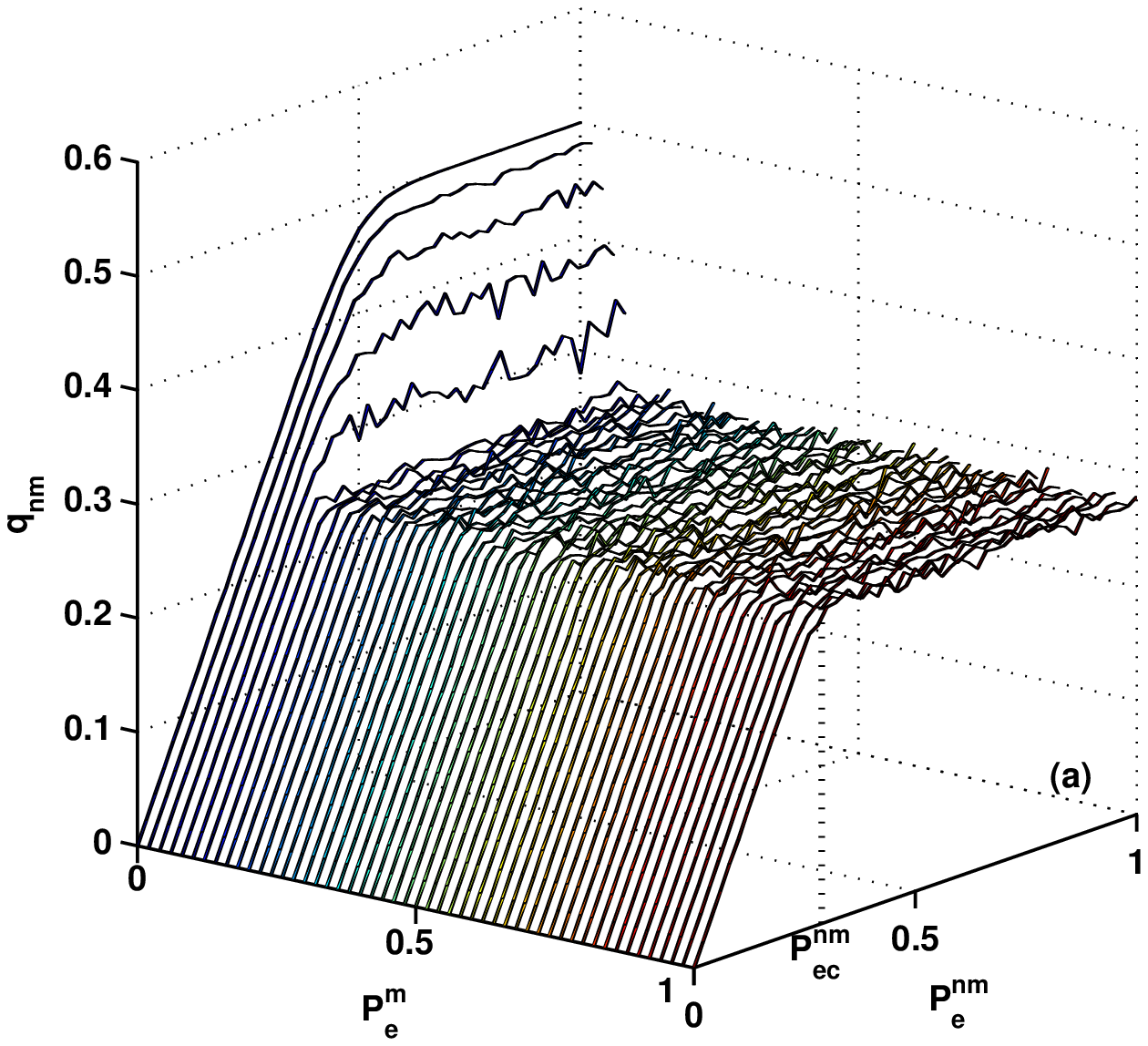}
\includegraphics[width=5 in]{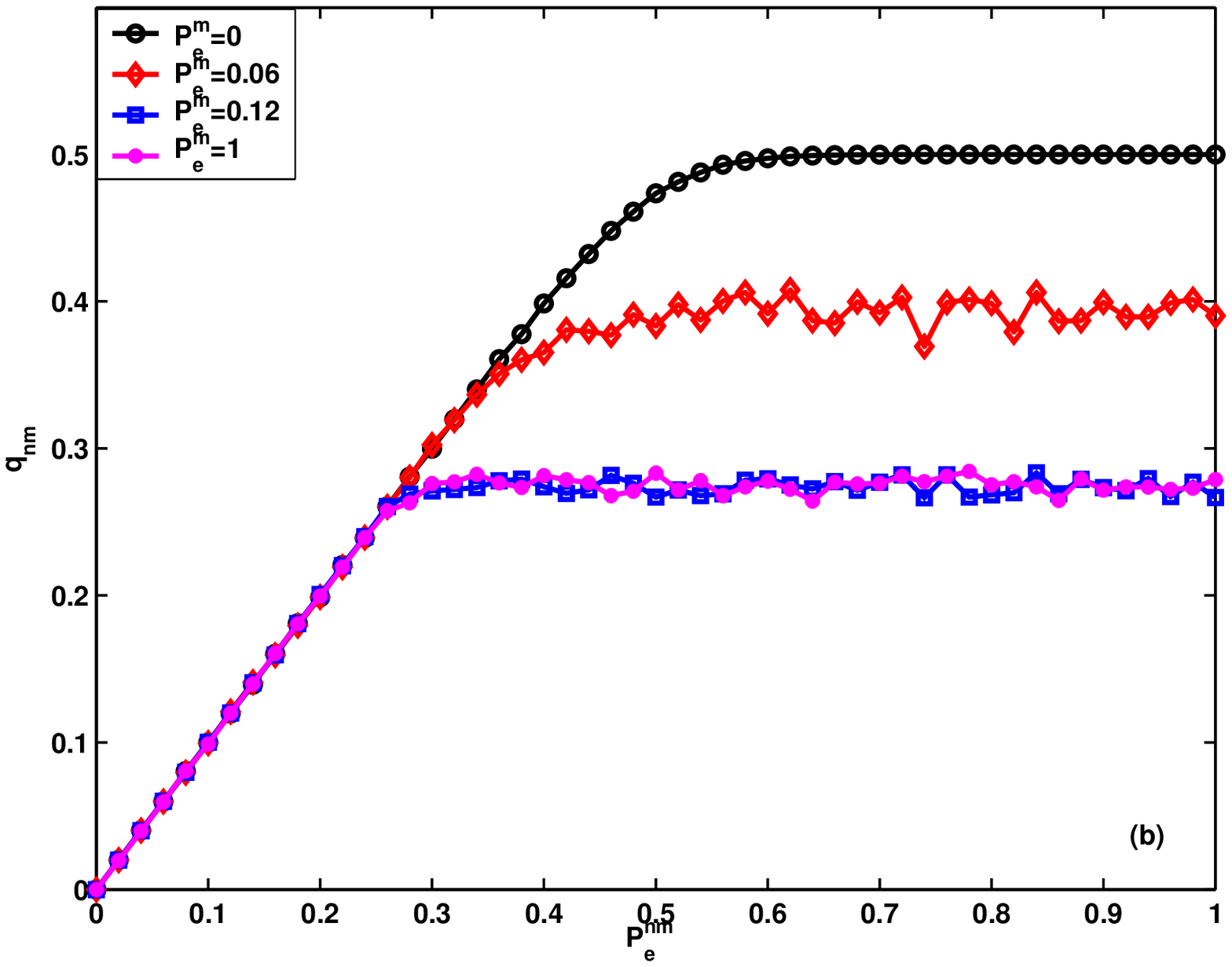}
  \caption{}\label{f3}
\end{figure}

\begin{figure}
  \includegraphics[width=5 in]{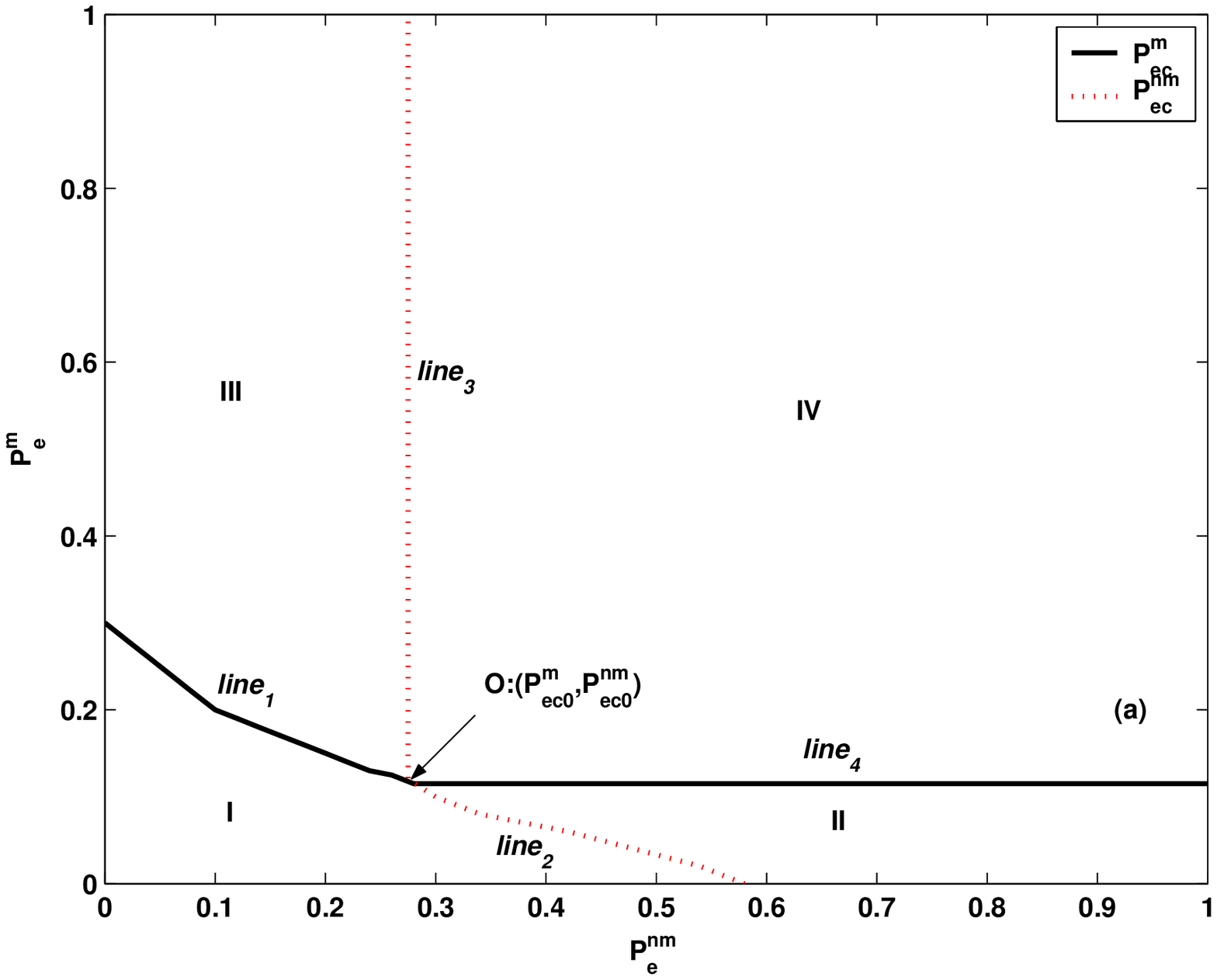}\\
   \label{f4a}
    \includegraphics[width=5 in]{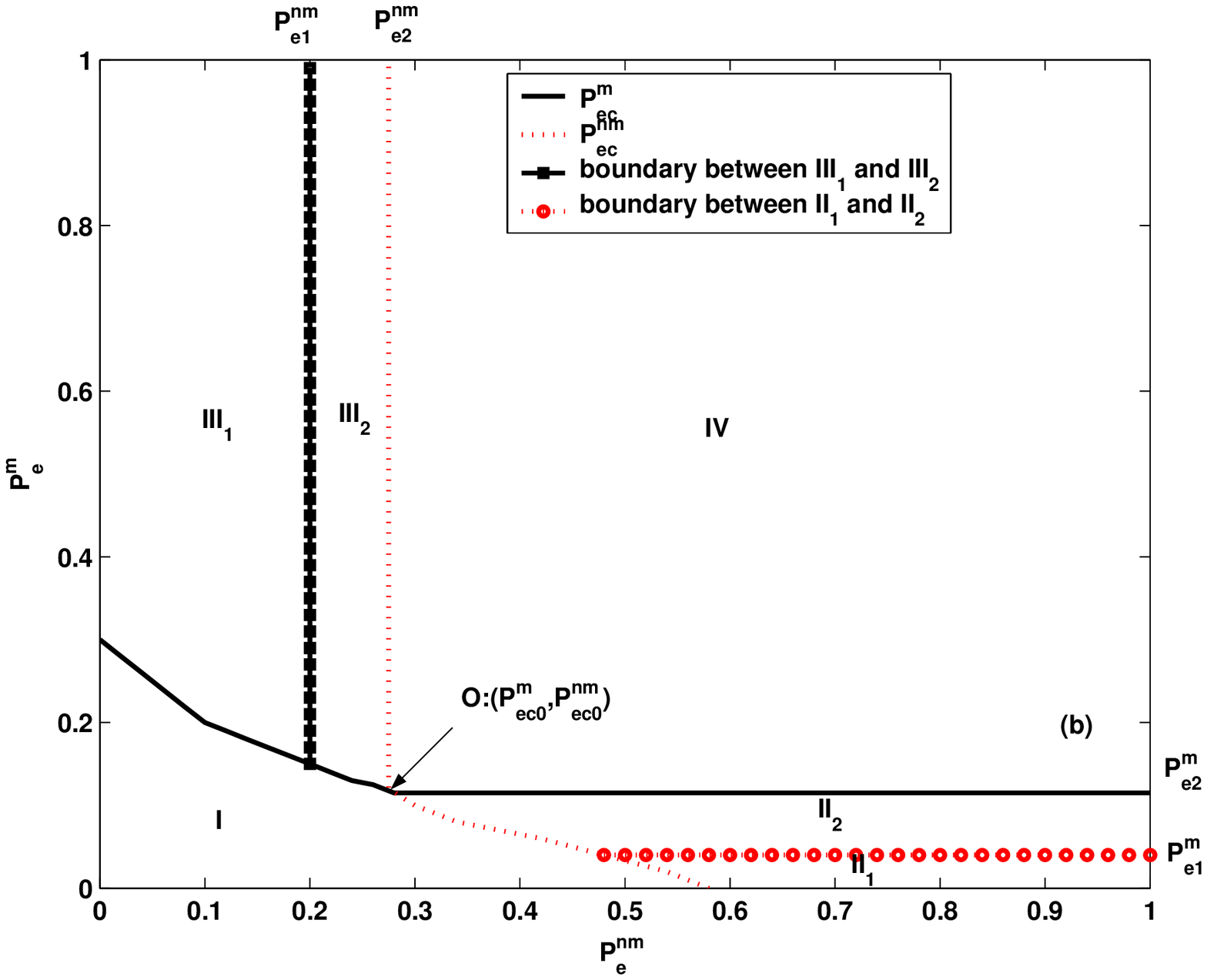}\\
  \caption{}\label{f4b}
\end{figure}

\begin{figure}
  \includegraphics[width=5 in]{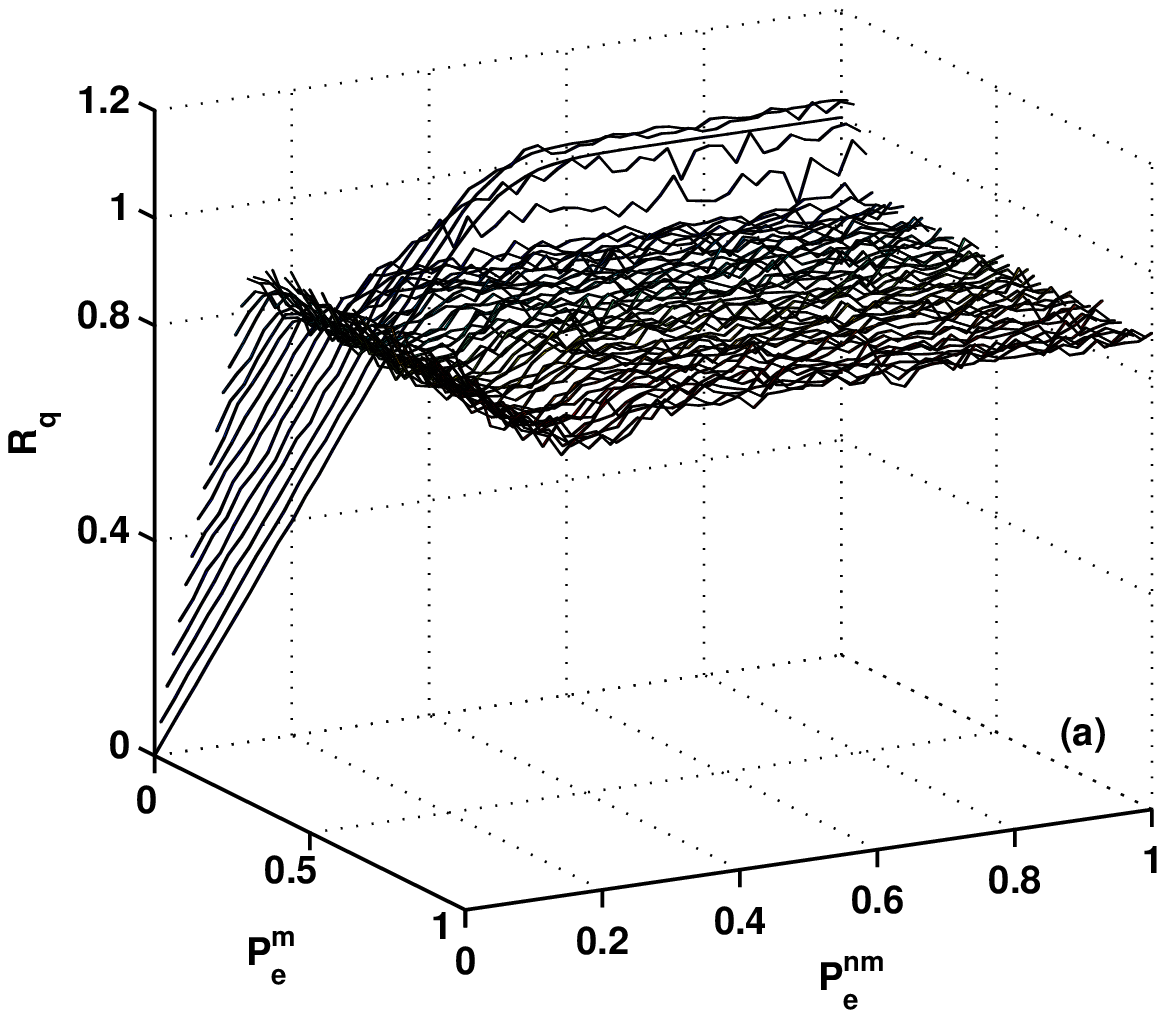}\\
   \label{f5a}

\end{figure}

\begin{figure}

  \includegraphics[width=5 in]{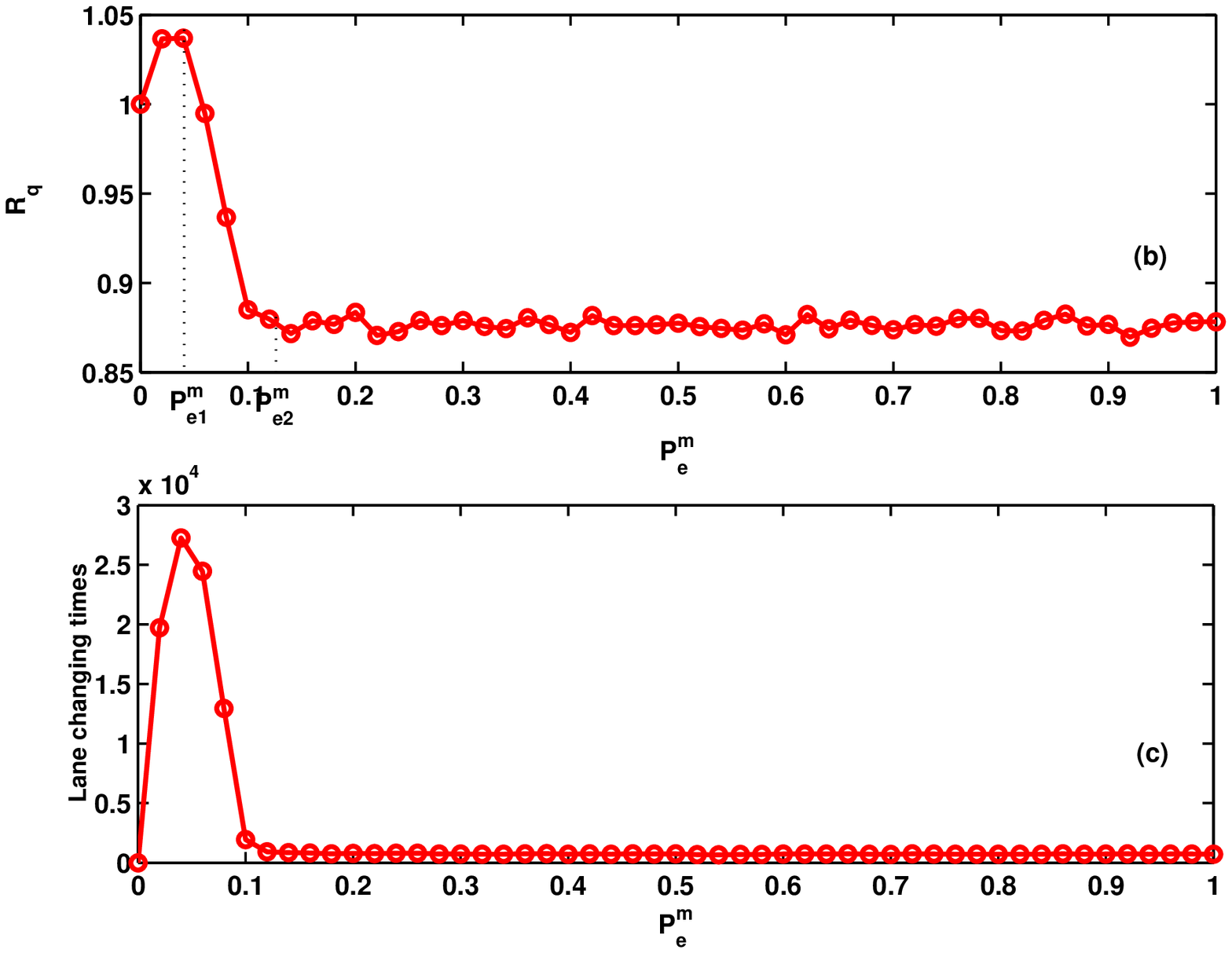}\\

    \includegraphics[width=5 in]{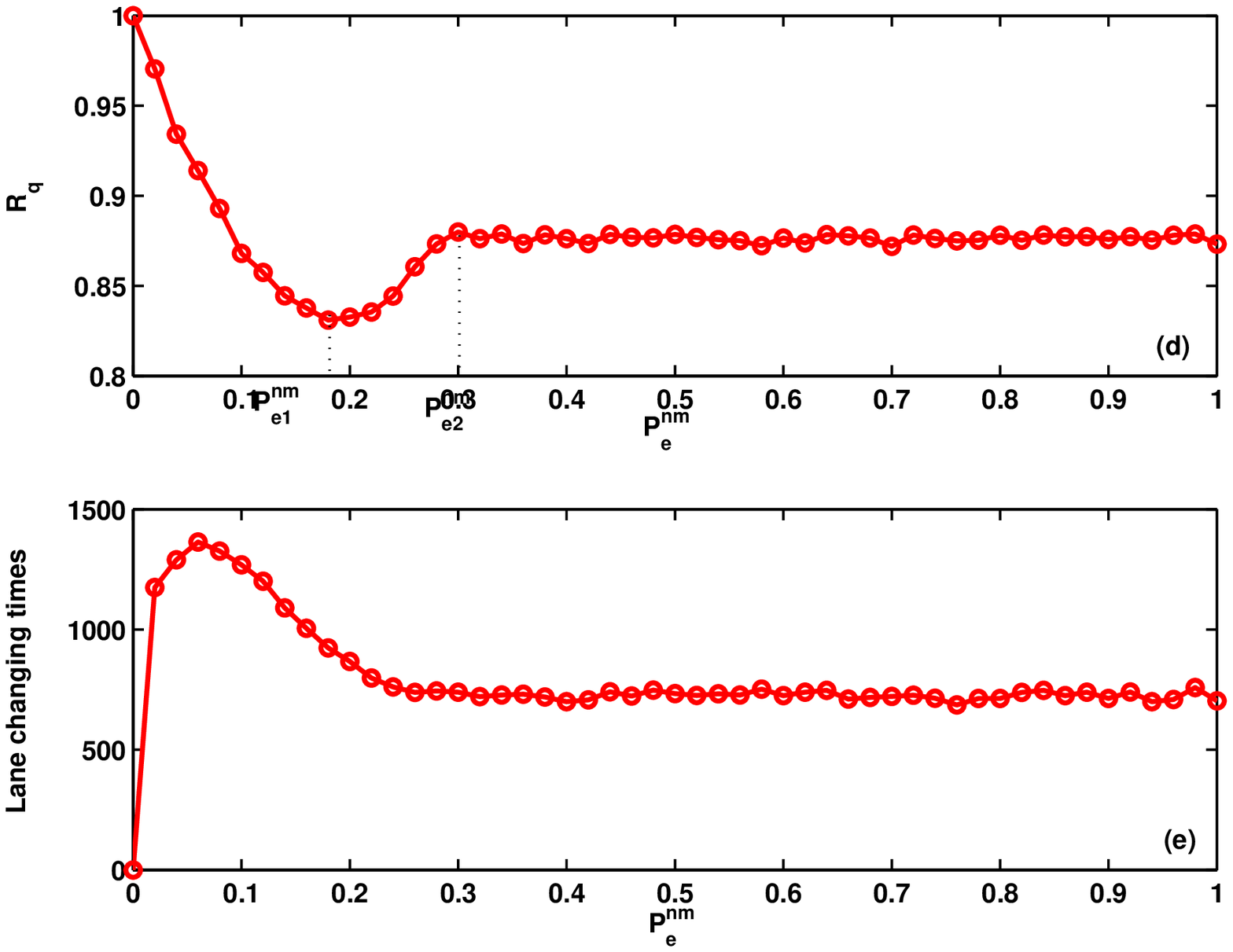}\\

  \caption{}\label{f6b}
\end{figure}

\end{document}